\documentclass[aps,prb,amsmath,amssymb,10pt,superscriptaddress,twocolumn,color,epsfig,graphicx,bm]{revtex4-1}

\usepackage{color}
\usepackage{graphicx}
\usepackage{graphics}
\usepackage{subfigure}
\usepackage{epsfig}
\usepackage{amsmath}
\usepackage{array}
\usepackage{amssymb}
\usepackage{graphicx}      
\usepackage{dcolumn}       
\usepackage{bm}            
\usepackage{array}
\usepackage{booktabs}

\usepackage[linktocpage=true,
  colorlinks=true, 
  pdfborder={0 0 0},
  linkcolor=blue,
  citecolor=red,
  filecolor=yellow,
  urlcolor=blue,
  bookmarks,
  pdfauthor={},
]{hyperref}

\newcommand{\tokyo}{Department of Applied Physics, The University of Tokyo, Tokyo 113-8656, Japan}

\begin{document}

\title{Unfolding energy spectra of multi-periodicity materials}

\author{Yu-ichiro Matsushita}   \affiliation{\tokyo}
\author{Hirofumi Nishi}   \affiliation{\tokyo}
\author{Jun-ichi Iwata}   \affiliation{\tokyo}
\author{Taichi Kosugi}   \affiliation{\tokyo}
\author{Atsushi Oshiyama}         \affiliation{\tokyo}

\date{\today}

\begin{abstract}
We propose a new unfolding scheme to analyze energy spectra of complex large-scale systems which are inherently of multi-periodicity. Considering twisted bilayer graphene (tBLG) as an example, we first show that the conventional unfolding scheme in the past using a single primitive-cell representation causes serious problems in analyses of the energy spectra. We then introduce our multi-space representation scheme in the unfolding method and clarify 
its validity for tBLG.  
Velocity renormalization of Dirac electrons in tBLG is elucidated in the present unfolding scheme. 
\end{abstract}

\pacs{~}
\maketitle

Understanding of the electronic band structures is prerequisite for understanding various material properties, such as the optical spectra, the direct or indirect band gaps, the electron transport, and so on. Electronic band structure calculations based on the density-functional theory (DFT)\cite{HK,KS} is one of the powerful tools to investigate those electronic properties. The information of the electronic band structure is also yielded by various experimental techniques: Hall measurements which yields the effective mass or the curvature of bands, photoluminescence measurements which yields the band gap. 
A more direct experimental method to observe the electronic band structure is the angle-resolved photo-emission spectroscopy (ARPES)\cite{ARPES}. The collaboration between DFT calculations and ARPES measurements is a promising approach to unveil the electronic properties. Recently this approach has been vastly used, and helped especially in deepening the surface science, exemplified by topological insulators\cite{ARPES,Topological1} and two-dimensional materials\cite{ARPES,graphene1,graphene2}.

In spite of the strong tool of DFT calculations collaborated with ARPES measurements, we often face a big issue of the {\it folded} bands when we use the supercell scheme. To discuss the effects of perturbations, such as some defects or interaction with substrates, which we call external fields hereafter, on the electronic structure of a periodic system, we routinely adopt a larger unitcell, called supercell, than the primitive unitcell of the system. However, using the supercell scheme, the size of the Brillouin zone (BZ) becomes smaller in reciprocal space and the electronic bands are folded into the smaller BZ. The folded bands result in a denser band structure than the original one, and the denser band structure is far from that observed by ARPES measurements. This makes it difficult to compare with the results of ARPES measurements directly.

To remedy this problem in supercell calculations, several {\it unfolding} techniques have been proposed on the basis of the tight-binding method \cite{Unfolding5,Unfolding6,Unfolding7,Unfolding8,Unfolding9} and the DFT\cite{Unfolding1,Unfolding2,Unfolding3,Unfolding4}. The ultimate goal of the band unfolding techniques is to get the one-body Green's function including the effect of an external field expressed as
\begin{eqnarray}
{\hat G}_{\rm unfold}^{-1}={\hat G}_0^{-1}-{\hat V}, 
\label{Dyson}
\end{eqnarray}
where ${\hat G}_{\rm unfold}$ represents the unfolded one-body Green's function, ${\hat V}$ does the external field, and ${\hat G}_0$ is the one-body Green's function of the system with no external fields. The conventional unfolding method starts from the one-body Green's function, ${\hat G}^{\rm SC}(z)$, of the supercell for a complex $z$:
\begin{eqnarray}
{\hat G}^{\rm SC}(z)=\sum_{n{\bf k}_{\rm SC}}\frac{\left|\Psi^{\rm SC}_{n{\bf k}_{\rm SC}}\right>\left<\Psi^{\rm SC}_{n{\bf k}_{\rm SC}}\right|}{z-\epsilon_{n{\bf k}_{\rm SC}}}, \nonumber
\end{eqnarray}
where $\left|\Psi^{\rm SC}_{n{\bf k}_{\rm SC}}\right>$ and $\epsilon_{n{\bf k}_{\rm SC}}$ is the $n$-th eigenstate and eigenvalue of the supercell system at the ${\bf k}_{\rm SC}$ point of the BZ. Then, the supercell Green's function is projected onto the BZ of the primitive cell 
as follows:
\begin{eqnarray}
{\hat G}_{\rm unfold}({\bf k},z)={\hat P}_{\bf k}^{\dagger}{\hat G}^{\rm  SC}(z){\hat P}_{\bf k}, \nonumber
\end{eqnarray}
where ${\hat P}_{\bf k}$ is a projection operator onto the ${\bf k}$ point of the primitive-cell BZ, i.e., ${\hat P}_{\bf k}=\sum_n \left|\Psi^{\rm PC}_{n{\bf k}}\right>\left<\Psi^{\rm PC}_{n{\bf k}}\right|$ with $\left| \Psi^{\rm PC}_{n{\bf k}} \right>$ being the $n$-th eigenstate of the primitive-cell (unperturbed) system at the ${\bf k}$ point. The spectral function at energy $\epsilon$, $A({\bf k},\epsilon)$, is defined as the trace of the unfolded Green's function,
\begin{eqnarray}
A({\bf k},\epsilon)=-\frac{1}{\pi}{\rm Im}\left[{\rm tr} \left({\hat G}_{\rm unfold}({\bf k},\epsilon+i\delta)\right)\right],
\label{spectr}
\end{eqnarray}
 where $\delta$ is a positive infinitesimal,
and finally the unfolded band structures are presented as the spectrum of Eq.~(\ref{spectr}). It is noteworthy that the spectral function has a periodicity of the adopted primitive-cell BZ, as is easily understood by considering that the projection operator ${\hat P}_{\bf k}$ has the translational symmetry of the primitive-cell BZ, ${\hat P}_{\bf k}={\hat P}_{{\bf k}+{\bf G}}$ with ${\bf G}$ being an arbitrary reciprocal lattice vector of the primitive cell. 

The conventional band unfolding techniques have been successfully used for many systems combined with the DFT calculations; For example, the direct comparison of the unfolded bands with experiments was done for Ni$_{1/3}$TiS$_2$ and marvelous agreement has been reported \cite{Matsushita}.


In the conventional band unfolding techniques, it is implicitly assumed that there is {\it single} primitive-cell BZ corresponding to the unperturbed system, and consequently the spectral function has "{\it one}" periodicity of the same primitive-cell BZ (single primitive-cell representation).
However, there exist the systems that the conventional unfolding techniques can not be applicable to. Two or more materials combined as a single system, such as two-dimensional materials on a substrate or stacks of different thin films, is a good example for such systems.
In such systems, they have two (or more) possible primitive-cell BZs corresponding to each periodicity of the constituent materials. With the conventional band unfolding method, we impose only {\it one} artificial periodicity of the adopted primitive cell, even though the system has two (or more) possible primitive-cell BZs. This causes serious artifacts that 
the unphysical ghost bands appear. In other words, the conventional unfolding scheme does not allow a reasonable unperturbed Green's function ${\hat G}_0$ for these systems. Here we call the system that the assumption of the single-primitive-cell representation does not hold the multi-periodicity system. 

In this paper, we present our detailed analyses of the serious problem associated with the conventional band unfolding method for multi-periodicity materials and introduce a new unfolding scheme to remedy the problem. For manifestation of the problem and demonstration of our successful new method, we consider a twisted-bilayer graphene (tBLG) as a target. tBLG is a sort of stacking materials composed of two slightly twisted graphene sheets via the van der Waals force, which exhibits two misfit primitive-cell BZs. In this study, we consider a tBLG with twist angle $\theta=9.43^\circ$ \cite{Uchida,Uchida2,Nishi}. 


\begin{figure}
\includegraphics[width=1.0\linewidth]{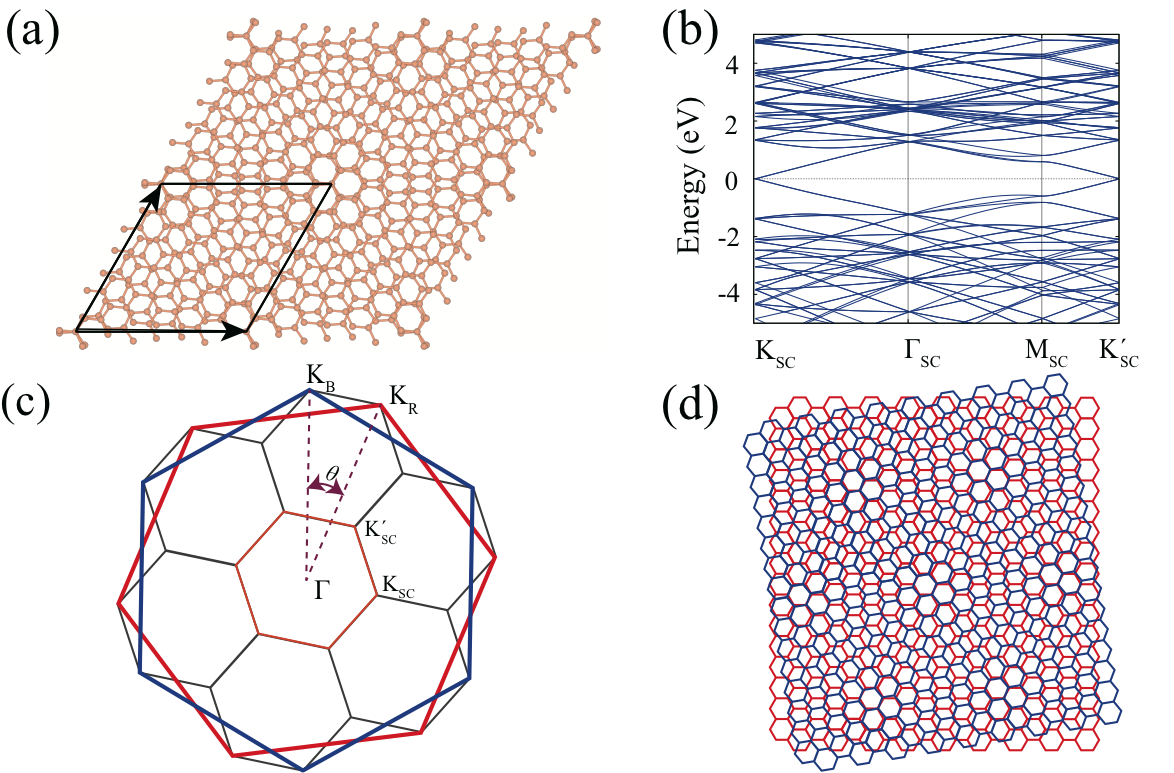}
\caption{(Color online). Atomic structure and BZs of tBLG. (a) Atomic structure
of tBLG with $\theta=9.43^\circ$ with black arrows being the lattice vectors of
the supercell. (b) Folded bands in the supercell BZ. (c) Two primitive-cell BZs of tBLG, where the red (blue) BZ depicts the primitive-cell BZ of the upper (lower) monolayer graphene. Black hexagons represent the supercell BZs. (d) The BZs
in the extended scheme for $\theta=9.43^\circ$.}
\label{Fig1}
\end{figure}

As an important feature of the tBLG, tBLG exhibits moir\'{e} pattern in real space as shown in Fig.~\ref{Fig1}(a), which inevitably makes us use the supercell scheme for the calculation of the system. In the supercell calculations, more than hundreds of atoms are inside the simulation cell, leading to the dense folded energy bands in the supercell BZ [See Fig.~\ref{Fig1}(b)]. More importantly, the primitive-cell BZ of each monolayer is twisted to each other by the same angle as in real space [Fig.~\ref{Fig1}(c)], and a moir\'{e} pattern emerges also in the reciprocal space [Fig.~\ref{Fig1}(d)]. Correspondingly, the spectral function 
does not have the periodicity of a primitive-cell BZ but that of the moir\'{e} pattern: i.e., The spectral functions represented in the 1st, 2nd, etc., primitive-cell BZs are different from each other. 


\begin{figure*}
\includegraphics[width=1.0\linewidth]{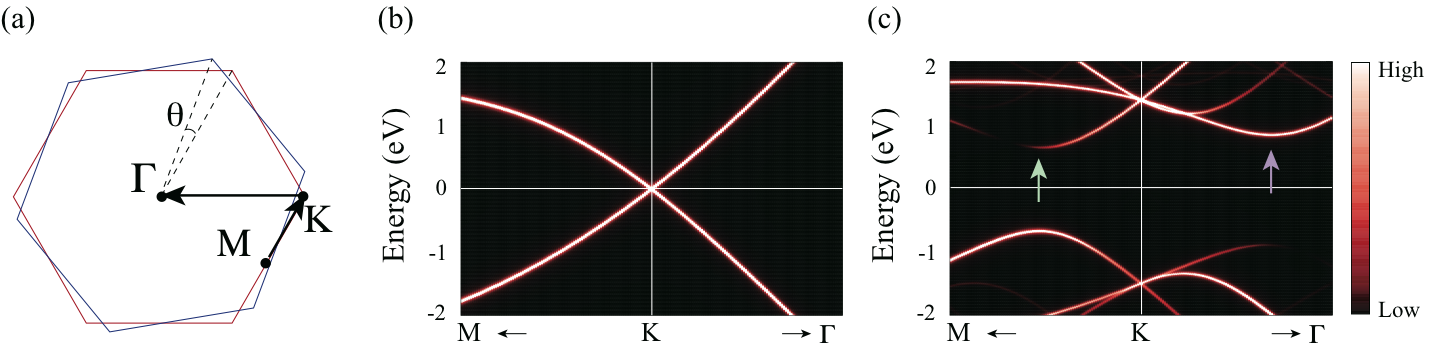}
\caption{(Color online). (a) Two primitive cell Blliouin zones corresponding to the upper layer (represented by the red hexagon), and the lower one (represented by the blue one). The unfolded band structure (calculated by the DFT) of the upper (b) and lower (c) monolayer graphene onto the red BZ shown in (a). Green and purple arrows indicate the ghost electronic bands.}
\label{Fig2}
\end{figure*}

To clearly manifest the problem, we first present the unfolded bands for the tBLG calculated by the DFT in Fig.~\ref{Fig2} with the conventional unfolding technique. We used an unfolding program code implemented in the Real-Space DFT program code (RSDFT)\cite{RSDFT1,RSDFT2,RSDFT3} for the calculations. In the RSDFT code, we introduce real-space grids and all the quantities including Kohn-Sham orbitals and thus valence electron densities are computed on the grid points. We used a local-density-approximation functional for the exchange-correlation energy\cite{PZ81}. The grid spacing has been taken to be 0.177 \AA\ corresponding to the 1060 eV cutoff in the plane-wave basis set. BZ integration has been performed with Monkhorst-Pack $5\times5\times1$ sample $k$ points. These parameters are adopted after the examination of the accuracy within 13 meV in total energy. The structural optimization has been done with a tolerance of  $2.6\times10^{-2}$ eV\AA$^{-1}$. The calculated energy bands for the "supercell" are unfolded to the primitive-cell BZ of the upper layer. 

If the interlayer interaction is switched off, the system should exhibit two clear periodicity in its spectral function. 
To clearly manifest that the conventional unfolding method does not reproduce the multi periodicity of band structure of the non-interacting tBLG, we have performed two independent unfolding calculations for each monolayer graphene. In these calculations, electronic bands of the upper (Fig.~\ref{Fig2}(b)) and lower (Fig.~\ref{Fig2}(c)) monolayer graphene are unfolded onto the upper layer primitive-cell BZ (the red hexagons in Fig.~\ref{Fig2}(a)). In Fig.~\ref{Fig2} (b), the unfolded band structure exhibits an unusual electronic states with linear dispersion at the $K$ point, which is a well known electronic state, called Dirac cone, in monolayer graphene. The intensity of the unfolded bands at the Fermi level is larger than those at the other $k$-points due to the degeneracy of the two electronic bands. As is important the most, the intensity of each electronic bands is uniform as what it should be, because there is no scatterers inside the monolayer system. In contrast, the unfolded bands in Fig.~\ref{Fig2}(c) exhibit non-uniform intensity (e.g., the electronic band pointed by the green arrow), as if they had some imperfectness causing finite lifetime. We have found that the reasons of these unphysical ghost bands is due to the curse introduced by imposing an artificial periodicity of the $k$-space to the system. Furthermore, the unfolded band structure repeats with the periodicity of the upper layer primitive-cell BZ, leading to no periodicity of moir\'{e} pattern in $k$-space anymore. This result clearly manifests that for multi-periodicity systems, bringing one periodicity via the projection in Eq.~(\ref{spectr}) constructs a wrong non-interacting Green's function, $G_0$, and causes artificial errors.


Then, a question arises here: How to fix the problem? We here naturally extend the definition of the spectral function. 
Our new definition of the spectral function should satisfy the condition that the correct multi-periodicity be recovered at the limit of the small external field, ${\hat V}\rightarrow 0$, in Eq.~(\ref{Dyson}), reflecting the fact that multi-periodicity materials have two or more primitive-cell BZs. We, then, decompose the projection operator ${\hat P}_{\bf k}$ in Eq.~(\ref{spectr}) to those of two subsystems of each consisting material, named (A) and (B), ${\hat P}_{\bf k}^{\rm (A)}$ and ${\hat P}_{\bf k}^{\rm (B)}$ (multi-space representation) as follows:
\begin{eqnarray}
{\hat P}_{\bf k}={\hat P}_{\bf k}^{\rm (A)}+{\hat P}_{\bf k}^{\rm (B)}.
\label{new_def}
\end{eqnarray}
The specific expression for the operators, ${\hat P}_{\bf k}^{(i)}$ ($i=A$ or $B$), is ${\hat P}_{\bf k}^{(i)}=\sum_n \left|\Psi^{{\rm PC }(i)}_{n{\bf k}}\right>\left<\Psi^{{\rm PC}(i)}_{n{\bf k}}\right|$, where $\left|\Psi^{{\rm PC }(i)}_{n{\bf k}}\right>$ represents the eigenstate of the primitive cell of each subsystem $i$.
In this present case of the tBLG, two subsystems correspond to the upper and lower monolayer graphene, respectively. 
By considering that each ${\hat P}_{\bf k}^{(i)}$ possesses a corresponding periodicity in $k$-space, ${\hat P}_{\bf k}^{(i)}={\hat P}_{{\bf k+G}^{(i)}}^{(i)}$ with ${\bf G}^{(i)}$ being the reciprocal lattice vector of the subsystem $i$. Therefore, the new unfolded bands maintain two periodicity coming from the two primitive-cell BZs and reproduce the correct non-interacting limit. In the practical construction of the projection operators, we can use the linear-combination-atomic-orbital (LCAO) basis set or the maximally localized Wannier functions for the complete basis set of the subspaces.

\begin{figure*}
\includegraphics[width=1.0\linewidth]{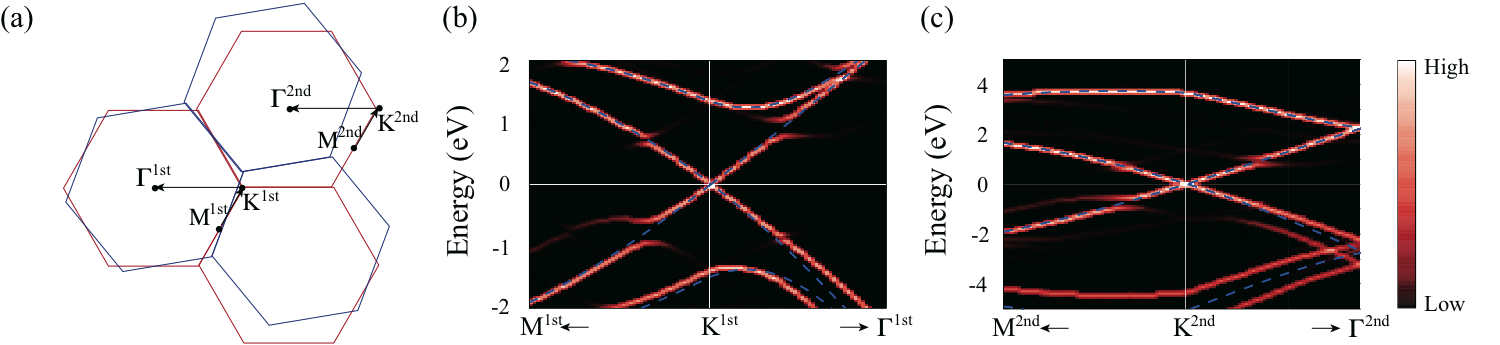}
\caption{(Color online). (a) Two primitive Blliouin zones corresponding to the upper layer (represented by the red hexagon) and the lower ones (represented by the blue ones). (b) The unfolded band structure (calculated by the tight binding method) along the symmetry line of the red 1st BZ as shown in (a). (c) The unfolded bands in the red 2nd BZ depicted in (a).}
\label{Fig3}
\end{figure*}

For demonstration, we applied the new unfolding method to the tBLG. First, for simplicity, we used the tight-binding model to verify our ideas. The tight-binding parameters we used are those in Ref.~\onlinecite{TB}. The results are shown in Fig.~\ref{Fig3}. Fig.~\ref{Fig3}(b) shows the unfolded bands in the 1st BZ along the pathway shown in Fig.~\ref{Fig3}(a). The overall behavior of the unfolded bands can be interpreted as the superposition of the two monolayer graphenes without layer-layer interaction as depicted by the blue broken lines in the figure. We can see that the unfolded bands show energy splitting at certain $k$ points. The location of the $k$ points are already known as originated from the cone-cone interaction in Ref.~\onlinecite{Nishi} and on the equidistant lines between adjacent two Dirac cones. Fig.~\ref{Fig3}(c) shows the unfolded band structures in the 2nd primitive BZ as drawn in Fig.~\ref{Fig3}(a). As clearly seen, the unfolded bands are completely different from that of Fig.~\ref{Fig3}(b) reflecting the misfit of the two primitive-cell BZs. This behavior is physically reasonable. In addition, surprisingly, compared with Fig.~\ref{Fig2}(c), it is clearly seen that the conventional band unfolding method yields artificial bands marked by the purple arrow in Fig.~\ref{Fig2}(c). The corresponding band does not appear in Fig.~\ref{Fig3}(b). 

Above, we have checked that our method works well for tBLG as an example of multi-periodicity matter with the tight-binding method. Next we applied the new unfolding method with full-DFT calculations. In this calculation, we used again RSDFT program code and applied it to the tBLG. The parameters we used are the same as those for the conventional unfolding calculations.
To build the localized wavefunction for the projection operators in Eq.~(\ref{new_def}), we adopted the simplest manner that we separate the real space unitcell into two subspaces at the middle of the two graphene layers so that each subspace includes one constituent. Then the projection operator of each subspace is constructed with the primitive-cell BZ of each constituent monolayer graphene. 

\begin{figure}
\includegraphics[width=1.0\linewidth]{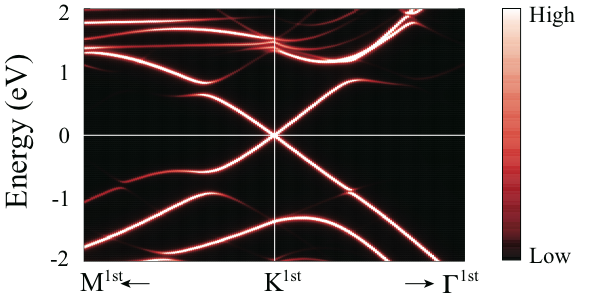}
\caption{(Color online). The unfolded band structure (calculated by the DFT) along the symmetry line of the red 1st BZ as shown in Fig.~\ref{Fig3}(a).}
\label{Fig4}
\end{figure}

The calculation results are shown in Fig.~\ref{Fig4}. Fig.~\ref{Fig4} shows the unfolded bands corresponding to the Fig.~\ref{Fig3}(b) with the new unfolding method. The overall features of the bands are in good agreement with those with the tight-binding method, Fig.~\ref{Fig3}(b). The unfolded bands exhibit energy splitting at the same $k$ points as those with the tight-binding method. 

In conclusion, 
we propose a new unfolding scheme to analyze energy spectra of complex large-scale systems which are inherently of multi-periodicity, such as two-dimensional materials on a substrate or stacks of different thin films. Considering twisted bilayer graphene (tBLG) as an example, we first show that the conventional unfolding scheme in the past using a single primitive-cell representation causes serious problems that ghost bands appear in the energy spectra. We then introduce our multi-space representation scheme in the unfolding method and clarify its validity for tBLG. Fermi velocity renormalization of Dirac electrons in tBLG is elucidated in the present unfolding scheme.






\begin{acknowledgments}
This research was supported by MEXT as ”Exploratory Challenge on Post-K computer” (Frontiers of Basic Science: Challenging the Limits). This research used computational resources of the K computer provided by the RIKEN Advanced Institute for Computational Science through the HPCI System Research project (Project ID:hp160265). Y.M. acknowledges the support from JSPS Grant-in-Aid for Young Scientists (B) (Grant Number 16K18075).
\end{acknowledgments}

\bibliographystyle{apsrev4-1}
\bibliography{paper}

\onecolumngrid


\end{document}